# Title: *In-Situ* Visualization of Long-Range Defect Interactions at the Edge of Melting

**Authors:** Leora E. Dresselhaus-Marais[1*], Grethe Winther[2], Marylesa Howard[3], Arnulfo Gonzalez[3], Sean R. Breckling[3], Can Yildirim[4], Philip K. Cook[5], Mustafacan Kutsal[6,7], Hugh Simons[7], Carsten Detlefs[6], Jon H. Eggert[1], Henning Friis Poulsen[7]

**Affiliations:**

[1]Lawrence Livermore National Laboratory, Physics Division, 7000 East Ave., Livermore, CA 94550.

[2]Technical University of Denmark, Department of Mechanical Engineering, Bldg. 425, 2800 Kgs. Lyngby, Denmark.

[3]Nevada National Security Site, 2621 Losee Road, North Las Vegas, NV 89030.

[4]CEA Grenoble, 17 Avenue des Martyrs, 38000 Grenoble, France.

[5]Universität für Bodenkultur Wien, Gregor-Mendel-Straße 33, 1180 Vienna, Austria.

[6]European Synchrotron Radiation Facility, 71 Avenue des Martyrs, 38000 Grenoble, France.

[7]Technical University of Denmark, Department of Physics, Bldg. 307, 2800 Kgs. Lyngby, Denmark.

*Correspondence to: Leora E. Dresselhaus-Marais dresselhausc1@llnl.gov

**One Sentence Summary:** A new tool reveals how patterns of microscopic defects coalesce and destabilize in bulk aluminum from 97-99% of the melt.

**Abstract:** Connecting a bulk material's microscopic defects to its macroscopic properties is an age-old problem in materials science. Long-range interactions between dislocations (line defects) are known to play a key role in how materials deform or melt, but we lack the tools to connect these dynamics to the macroscopic properties. We introduce time-resolved dark-field X-ray microscopy to directly visualize how dislocations move and interact over hundreds of micrometers, deep inside bulk aluminum. With real-time movies, we reveal the thermally-activated motion and interactions of dislocations that comprise a boundary, and show how weakened binding forces inhomogeneously destabilize the structure at 99% of the melting temperature. Connecting dynamics of the microstructure to its stability, we provide important opportunities to guide and validate multiscale models that are yet untested.

**Main Text:**

## Introduction

Defects underlie many of the mechanical, thermal, and electronic properties of materials. A prominent example is the dislocation, which is an extended linear defect in the atomic lattice that enables crystalline materials to permanently change their shape under mechanical loading[a]. The remarkable range of hardness and workability in ductile materials occurs because of how their dislocations can move and interact. Dislocations have been characterized extensively since the advent of transmission electron microscopy (TEM) (*1–5*), however, TEM requires sub-micrometer sample dimensions that introduce size effects and surface stresses that are not representative of bulk materials (*6*). *In-situ* TEM has thus been limited to resolving the dynamics of dislocation interactions in spatially localized processes (*7*) including the dynamics of isolated dislocations (*8*, *9*), the operation of dislocation sources (*10*) and interactions between dislocations and grain boundaries (*11*). To understand how defects dictate a material's macroscopic properties, we must resolve non-localized and stochastic processes. Under those conditions, defect motion is governed by unpredictable inhomogeneities in the sample that span a wide range of length-scales in 3D. Simulations have aided our understanding of such stochastic dislocation dynamics (*12–14*), but models of realistic patterns that span the necessary nanometer to millimeter length-scales have proven elusive. At this time, the specific interactions that cause a population of free dislocations to order into a structure (*polygonise*) are limited to theory and *ex-situ* studies (*15*), which cannot capture the dynamics. To understand how dislocations in bulk materials pattern in three-dimensional structures, we require new experimental tools.

[a] Dislocations move along the close-packed layers (termed glide planes) of the crystal, causing layers of atoms to slide past each other and thus induce permanent shear deformation (*6*). The Burgers vector is defined as the direction

The limitations of our measurement technology are especially clear at temperatures on the verge of melting. The mechanism by which ordered solids melt into disordered liquids at equilibrium has been actively contested for over a century (*16*, *17*). The criteria for equilibrium melting were first defined by the Lindemann and Born models, which describe melting as lattice destabilization from high-amplitude vibrational waves (*18*) or a "rigidity catastrophe" from loss of shear strength (*19*). Over the years, theory and experiments have studied the validity of these theories and have recently connected them based on contributions from the microstructure (*20–22*). Dislocations have been predicted to play a key role in seeding and nucleating melting (*23*, *24*), however, models still lack experimental guidance to determine the relevant physics for the exotic dislocation interactions in conditions at the cusp of melting. New advances in melt theory require experiments to inform the relevant physics over the necessary real-world conditions.

X-ray diffraction-based imaging has become a promising candidate for the necessary multi-scale characterization tools, as it can map crystallographic properties in the bulk and in 3D. While several X-ray methods have been able to resolve sub-surface dislocations, at this time, they have been unable to measure dynamics with sufficient resolution in space and time over a representative region. Topography (*25*) and topo-tomography (*26*) have been used to image dislocations for decades, but are limited to spatial resolution in the μm range. New X-ray nanobeam studies can achieve much higher spatial resolution, but require rastered scans that cannot probe dynamics over the necessary sub-second timescales (*27*). Dark-field X-ray microscopy (DFXM), analogous to its TEM counterpart, was recently developed to directly map the subtle deformations surrounding defects and boundaries beneath a material's surface—giving new views of the microstructure (*28*). While DFXM has addressed key issues in ferroelectrics (*29*) and biominerals (*30*), it has only recently been applied to dislocation studies (*31*). For materials with sufficiently low dislocation densities, DFXM was demonstrated to resolve single dislocations by mapping the strain fields around dislocation cores (weak-beam contrast) (*31*, *32*).

of the shear imparted by a dislocation; the crystallographic lattice around dislocations is distorted by elastic displacement fields (strain) that changes the energy of nearby bonds—creating force fields (stress) that allow them to interact. Dislocations are characterized as having edge, screw or mixed edge/screw character depending on the geometric relationship between the Burgers vector and the dislocation line.

We present time-resolved DFXM as a new tool to map how dislocations move and interact in delocalized processes deep inside bulk materials. With this approach, we resolve the individual and collective motion of the dislocations in a dislocation boundary (DB) ~200-µm beneath the surface of single-crystal aluminum. Our images map how the DB migrates along a very low-angle boundary (LAB) as it is heated from 97% to 99% of the melting temperature, $T_m$ = 660 °C. We zoom-in on how dislocations enter and leave the boundary, causing two DB segments to coalesce and stabilize into one cohesive structure. As the DB subsequently migrates and increases its spacing between dislocations, we observe how the boundary destabilizes. Connecting this to theory, we reveal the mechanism by which the dislocation boundary dissolves at the cusp of melting, as thermal forces dominate dislocation interactions. By visualizing and quantifying thermally-activated dynamics that were previously limited to theory, we demonstrate a new class of bulk measurements that is now accessible with time-resolved DFXM, offering key opportunities across materials science.

## Results & Discussion

We use *in-situ* DFXM to resolve how an ensemble of dislocations evolve in an aluminum crystal as we slowly heat it, recovering it towards a pristine non-defected form (*thermal annealing*). As shown in Fig. 1a, the X-rays illuminate a single *observation plane* in the sample, allowing the microscope to map the local structure over a 200 x 400 x 0.6 µm³ internal region with ~300-nm resolution and 250 ms between frames. To minimize effects besides temperature, we apply no external mechanical stress and use a high-purity sample to avoid competing solute-dislocation effects. We present the evolution of structure over 12.5 minutes, collecting scans of ~120-s at each temperature as we heat the crystal from 0.97 to 0.99 $T_m$ in 2 °C increments. Each temperature increase occurs over short 3-10 s intervals between movies (see Supplement for full thermal history). These controls ensure that the dislocation motion we observe arises from the interaction forces between neighboring dislocations, vacancy concentrations and thermal expansions from local temperature fluctuations.

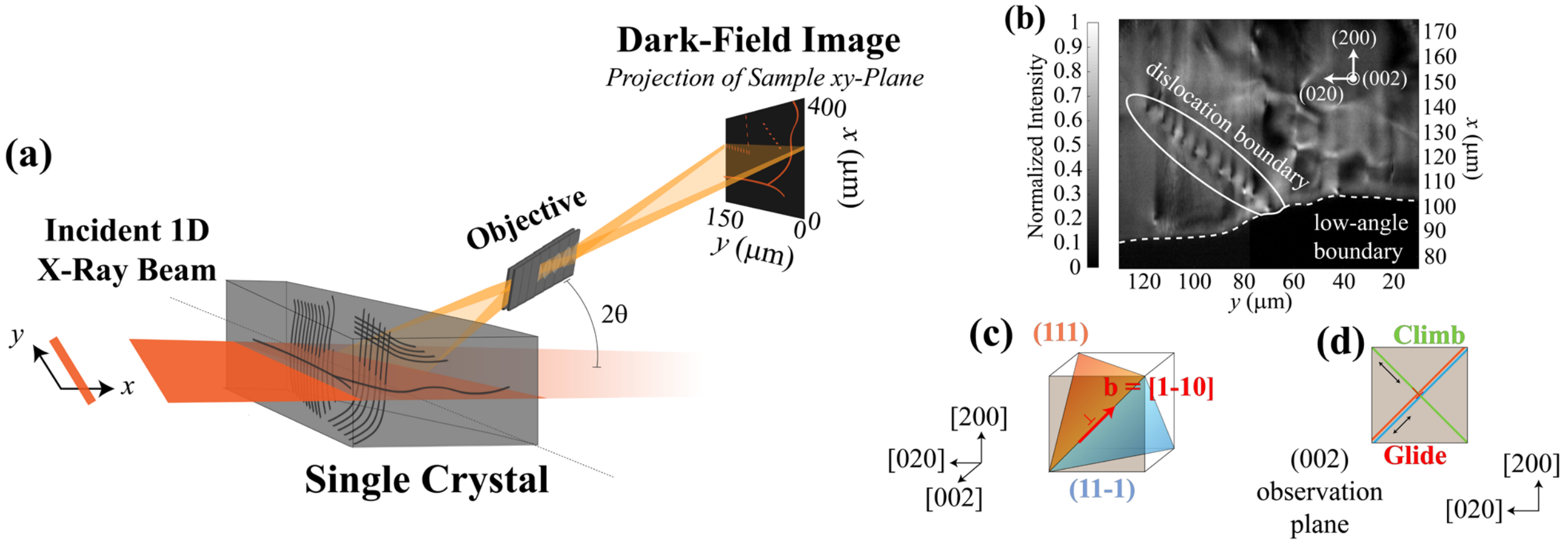

**Figure 1.** (a) Schematic of the experimental geometry for the dark-field X-ray microscope. We show the Single Crystal with free dislocations and their arrays inside a single crystal to demonstrate how the X-rays illuminate a 600-µm thick *observation plane* that projects and magnifies the local structure onto the detector. (b) A zoomed-in image of the single domain we focus on for this study, which includes a collection of dislocations (bright spots) at T = 638 °C (0.98$T_m$). The white oval outlines an array of dislocations that form a dislocation boundary, which is pinned to a low-angle boundary (dotted white line). We show a schematic (c) of the Burgers vector, **b**, (red) and slip planes (blue and orange) of the edge dislocations in the tilt boundary and (d) show how the climb and glide directions for these dislocations trace through the observation plane.

The bulk single-crystal in this study forms a large internal pristine domain (Fig. 1b) during the annealing treatment, which is surrounded by very low-angle grain boundaries (LABs). As shown in Fig. 1b, the crystalline domain includes a population of individual free dislocations and collective structures. The dashed white line identifies a LAB that a full scan at lower temperatures confirms is misoriented ~0.01° with respect to the primary domain in the image (details in Supplement). Above the LAB, several alternating bright-dark regions correspond to the long-range deformation fields surrounding individual dislocations. In this case, the dislocation lines slice through the observation plane with a steep incline, as illustrated for an analogous set of dislocation arrays in Fig. 1a. Based on the motion of these dislocations, we identify them as edge dislocations with identical $[1\bar{1}0]$ Burgers vectors – lying in the observation plane (details in Supplement S1). Our interpretation is supported by the predicted strain and rotation fields, which are consistent with simulations of the raw images we collect (*32*). Circled in white in Fig. 1b, the evenly spaced array of dislocations in the DB packs along the trace of their glide plane in our images (45° from [020]). The geometry agrees with the DB being a tilt boundary that is packed along a plane normal to the Burgers vector. We show the geometry of the Burgers vector and the slip planes for the boundary dislocations in Fig. 1c, projecting the structure onto our (002) observation plane to define the directions that correspond to glide and climb mechanisms of dislocation motion in our images (Fig. 1d).

Tilt boundaries of this kind are known to form a stabilizing dislocation structure, as their adjacent strain fields counter-act each other, reducing the strain energy between neighboring dislocations in a process called *stress screening* (*6*).

## Snapshots of the Progress of DB Evolution

We begin this study by showing snapshots that resolve a "static picture" of how the boundary evolves over the temperature range. Even single snapshots from each of the 8 movies reveal a new view of long-range DB motion. Fig. 2a includes an image of the crystalline domain with the positions of all DB dislocations in one frame for each temperature. With this view, we see that the dislocations are spaced ~5-9 µm apart in this temperature range, which is ~10x larger than previous observations in deformed metals (*3*, *5*). The wide spacing is not unreasonable for this recovered crystal at near-melt temperatures and demonstrates that stress screening still stabilizes a DB even with weak forces, in the absence of competing interactions.

Fig. 2a also reveals that the DB's trace migrates as the temperature increases, moving a total of ~60-µm until it shifts out of the field of view. The DB shifts along the LAB, suggesting an interaction between the two structures as the number of dislocations in the boundary decreases. Fig. 2a also shows that the DB begins as two separate boundary segments (Fig. 2b) that coalesce at 638 °C to form a single segment (Fig. 2c). A qualitative look at the angle between [020] and the DB trace demonstrates that the lower segment initially has an unfavorable 19° trace that rotates towards the more favorable 45° as it closes the distance between the two DB segments.

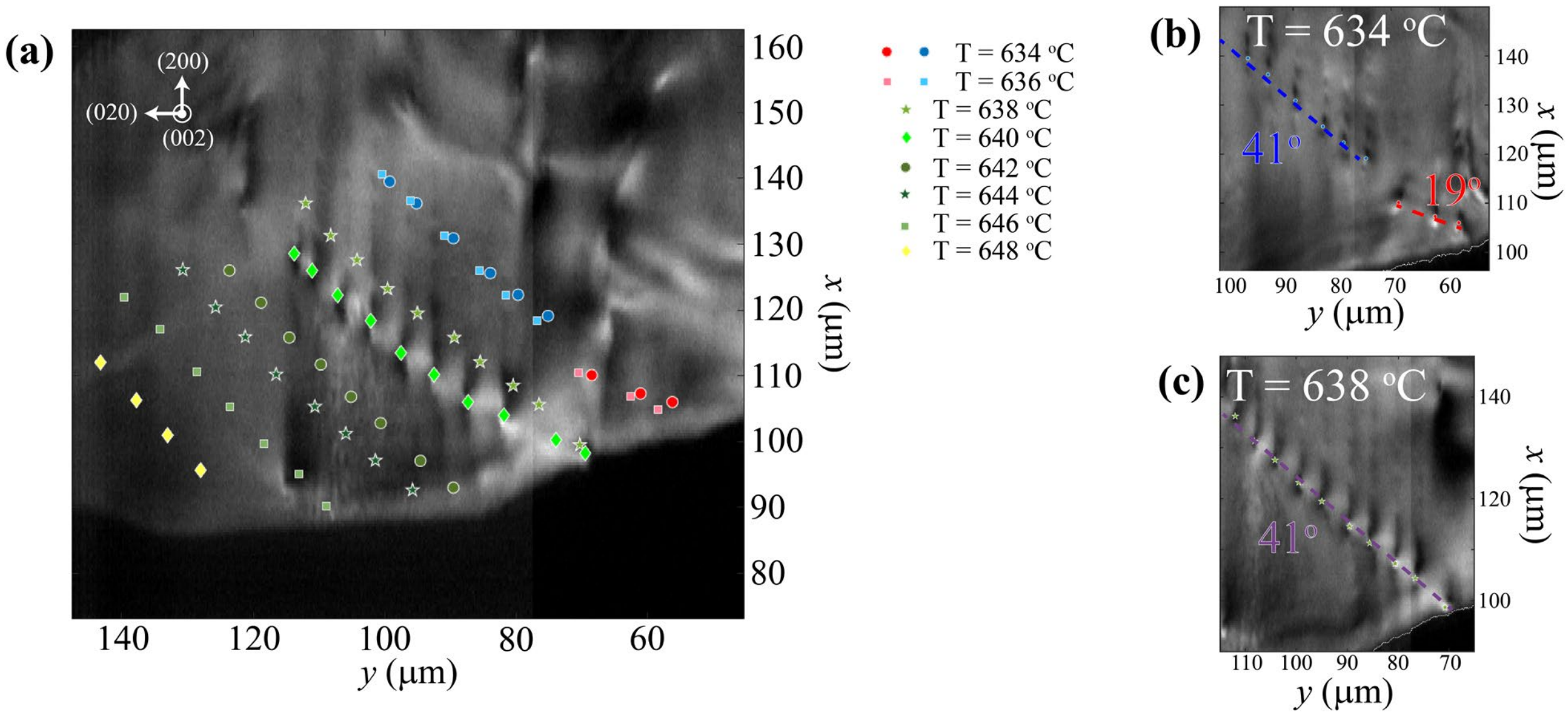

**Figure 2.** (a) The motion of the dislocation boundary shown in Fig. 1b with increasing temperature from 634 °C (0.97$T_m$) to 648 °C (0.99$T_m$), shown by the interpolated positions of all DB dislocation cores, plotted over an image from T = 640 °C. We zoom in on the DB at (b) T = 634 °C and at (c) T = 638 °C to show the joining of the two DB segments. The lines are best fits to straight boundaries. Images and associated linear fits are provided in the Supplement.

With only the snapshots in Fig. 2, we cannot resolve how the DB segments join, which dislocations disappear and how the spacing between dislocations increases as the DB migrates along the LAB.

## Coalescence Stabilizes the DB Structures

To resolve how the two DB segments coalesce, we turn to the real-time movie at 638 °C (Supplementary Movie 1). While Fig. 2a-c indicates that the lower segment rotates towards the upper one before they join, the time-resolved view from representative frames in Fig. 3a-d reveals that a lone dislocation (D3) inserting into the lower boundary ultimately drives the dislocations to redistribute. Using computer-vision tools with manual corrections to enhance the precision (*33*), we track the position of each dislocation over the full insertion mechanism, plotting the components along the climb and glide directions that correspond to the different types of dislocation motion (Fig. 3e-f). The plots and images in Fig. 3 demonstrate that the DB coalescence includes contributions from the four dislocations nearest to the LAB (labeled D2-D5). In Fig. 3 we plot these with D6 to correlate the insertion mechanism to motion in the remaining DB.

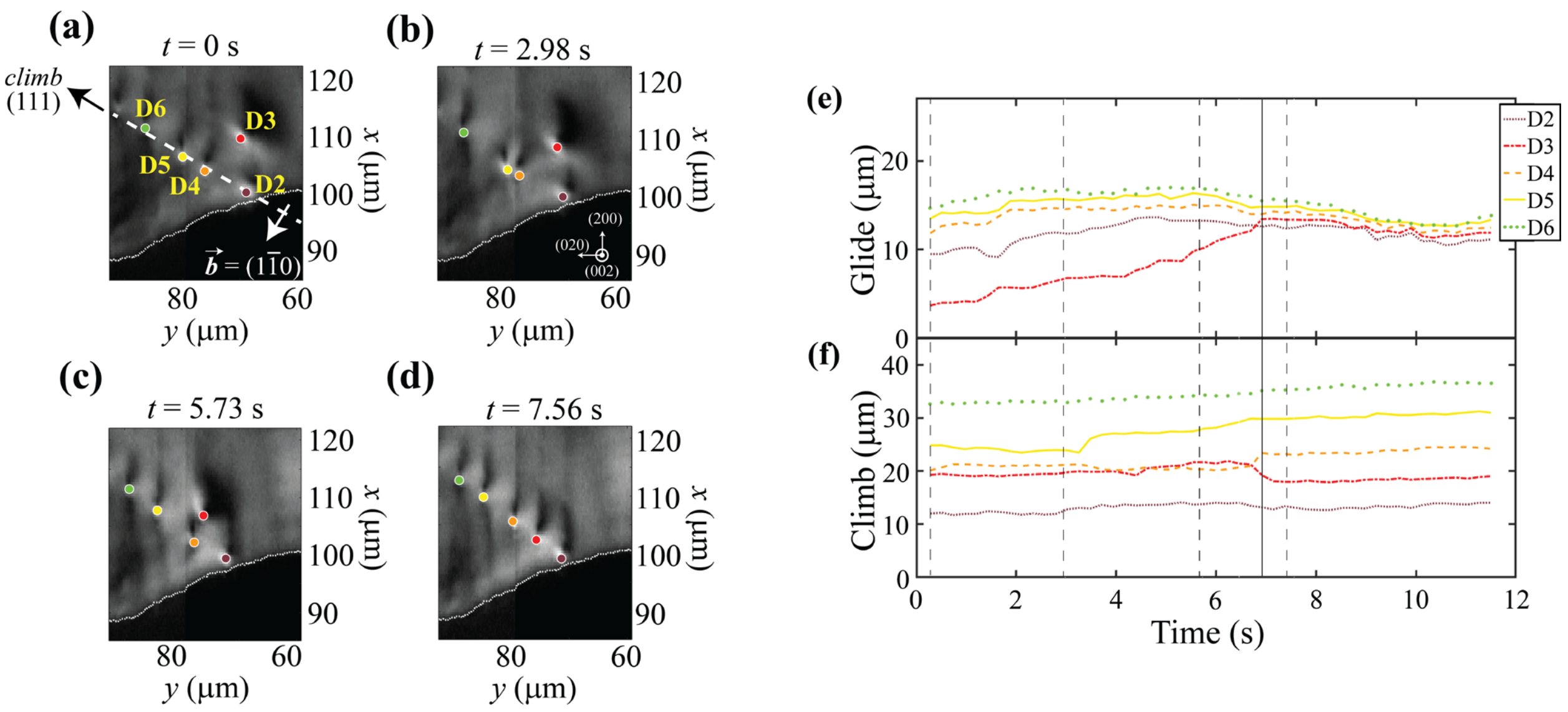

**Figure 3.** (a-d) Four representative frames showing how dislocation D3 inserts into the lower DB segment at T = 638 °C. We label the active dislocations in the first frame (a) and plot them for all subsequent frames to emphasize each dislocation's position. We also include plots of the position of each annotated dislocation, projected onto the (e) glide and (f) climb directions (as indicated in (a)). The vertical dotted lines mark the times of each frame in (a)-(d) and the solid vertical line marks the time at which D3 enters the DB.

As shown by the frames in Fig. 3a-b and the trace in Fig. 3e, the free dislocation, D3, initially glides towards the boundary, while D4 and D5 climb gradually towards each other. After ~3 s, D4 and D5 reach a critical distance and repel each other, displacing D4 into a pile-up geometry with D3. While the attractive climb between D4 and D5 is slow, their repulsion is faster than our 250-ms time resolution. The newly formed pile-up geometry theoretically induces a repulsive force between D4 and D3 along the glide plane (Fig. 3c), which should require D3 to move via climb to insert into the boundary. We resolve that the deformation fields for D2 and D3 overlap as D3 migrates further toward the boundary, suggesting that D2 and D3 interact to facilitate the climb. Passing <300 nm from D4, D3 ultimately climbs into the boundary, changing its shape as it inserts. D3's shape-change indicates that the deformation field changes upon insertion. Following these interactions, all five dislocations slowly migrate along the climb direction until they settle into their most favorable positions—closing the gap between the two segments.

The insertion mechanism and its associated coalescence of the two DB segments illustrates the stabilizing character of the tilt boundary.

## Dynamics over the Full Temperature Range

Our observations from Fig. 2a of the reduced number of dislocations and increase in spacing between DB dislocations can also be resolved more clearly with time-resolved measurements. We resolve the pathway by which dislocations leave the boundary by tracking the positions of all dislocations in the DB from T = 636-646 °C. For the full 12.5-minute acquisition, we project the components of each position along the glide and climb directions over each temperature, as shown in Fig. 4a and b respectively. The black bars indicate the time over which the temperature increases between each acquisition. Changes to the spacing between neighboring dislocations are clearest along the climb direction—where motion requires vacancy diffusion—while the change of the trace angle and the DB migration are clearest along the glide direction.

The full traces in Fig. 4a-b reveal how six dislocations leave the DB. The brown triangles in Fig. 4a-b mark five positions where dislocations exit the boundary via absorption into the LAB, while the brown stars mark the position where one dislocation, D11, escapes the boundary by slowly migrating via climb into the interior of the crystalline domain, likely due to other dislocations that are nearby (see Supplementary Movie 2). Representative frames from T = 640 °C are shown for the absorption and escape in Fig. 4c and 4d, respectively, with the exiting dislocations circled in red for clarity. The relative abundance of dislocations exiting the DB at the LAB reveals the importance of the LAB in increasing the spacing between boundary dislocations. Looking more closely at the LAB highlights this point. Dislocations are absorbed by first climbing towards the LAB until they affix to it; the remainder of the DB then glides past the immobile dislocation as the junction dislocation slowly exits the crystalline domain and moves out of our field of view. This mechanism indicates that the LAB can impose stronger interaction forces than those that stabilize the DB, however, these forces only act over much smaller distances. The abundance of dislocations that absorb into the LAB defines it as a dislocation *sink*, as seen in TEM (*34*). The successive absorption events in Fig. 4 demonstrate the first direct view of how individual dislocations absorb into a sink in *bulk* metals.

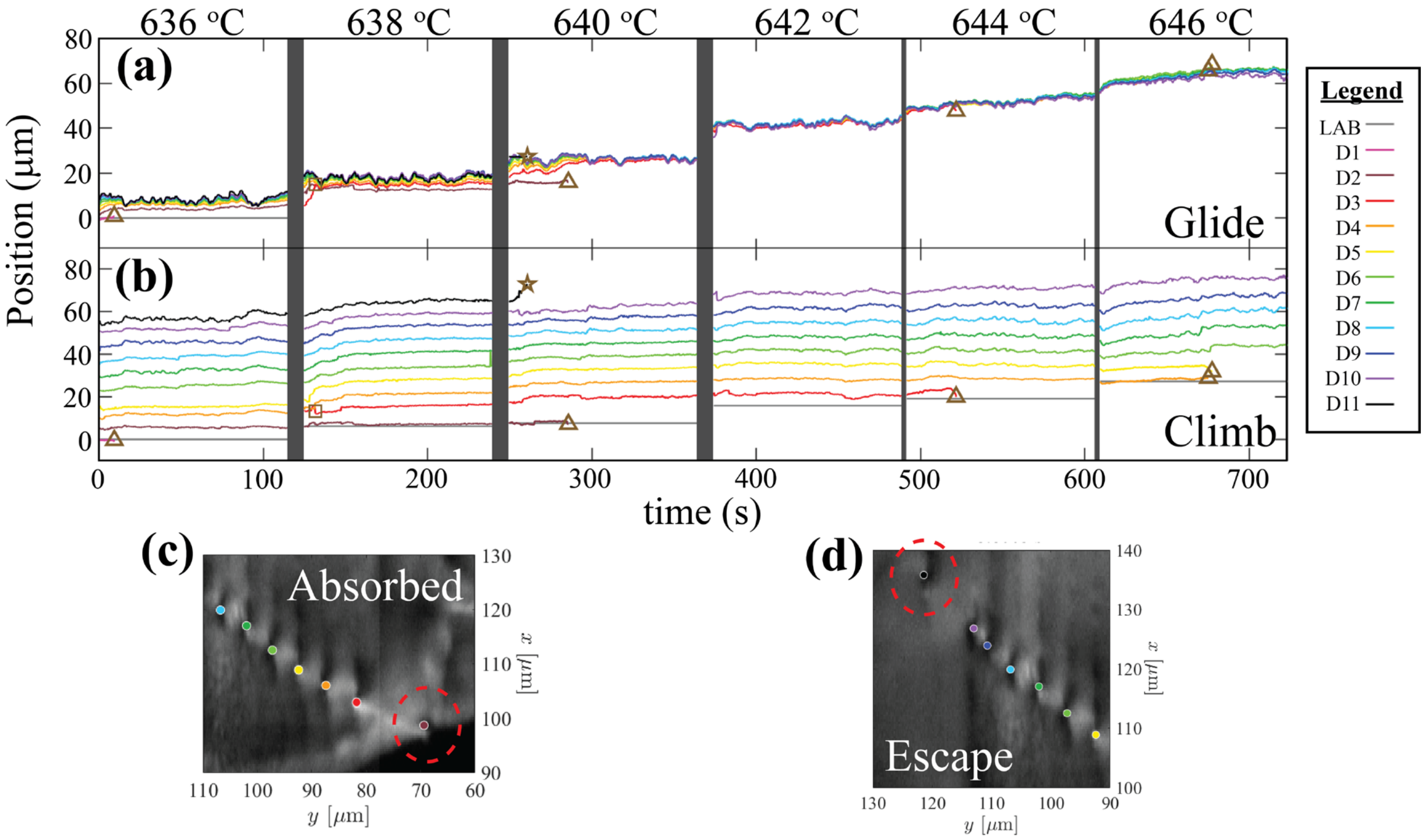

**Figure 4.** Plots showing the full evolution of the DB over six temperatures, from 0.97–0.99$T_m$. We show the position of each dislocation in the boundary at each time and resolve the motion along the glide (a) and climb (b) directions. In both plots, 0 corresponds to the position of the first dislocation in the first frame at T = 636 °C. The positions where dislocations are absorbed into the LAB are marked by brown triangles, the position where a dislocation escapes into the crystalline domain is marked by a star, and the position at which D3 inserts into the DB is marked by a square. To demonstrate how dislocations exit the boundary, we show representative frames of (c) how D2 is absorbed into the LAB, and (d) how D11 escapes into the crystal, both at T = 640 °C (exiting dislocations are circled in red).

## A Closer Look at DB Stability as $T \rightarrow T_m$

Beyond dislocation interactions, Fig. 4a-b also demonstrates collective dislocation motion in the DB with subtle trends at the highest temperatures that reveal key changes at these near-melt conditions. After the DB orients along its preferred 45° trace ($t > 300$ s), the dislocations oscillate collectively along both the climb and glide directions via thermally-activated motion (Supplementary Movie 2). Thermal motion is a stochastic process that arises from local inhomogeneities in the sample; climb motion is driven by vacancy diffusion, while glide motion is driven by the local stresses imposed by long-range dislocation interactions (*35*). At the verge of melting, thermal motion of lattice defects differs from predictions in classical dislocation theory. Lindemann theory describes melting as a vibrational catastrophe that destabilizes lattice bonds when the thermally amplified atomic vibrations reach the critical temperature, $T_m$ (*18*, *19*). Larger length-scales resolve the Lindemann criteria as increased dislocation mobility as $T \rightarrow T_m$, as the additional disorder creates new pathways that facilitate their motion (*22*).

Our experiments resolve the increased dislocation mobility as random variation in the position of each dislocation between frames, which we quantify as the temporal variance in dislocation position. We corroborate that variance relates to mobility, as the variance increases just before each dislocation is expelled from the boundary, then reduces after the dislocations leave (Fig. S4 in Supplement). In the absence of dislocation expulsions, however, we see an increase in dislocation mobility at the highest temperatures. Fig. 5a plots the average variance in the position of boundary dislocations for each temperature, computed from dislocation motion over the full timescan at each temperature from T = 642-646 °C. The variance in dislocation position jumps by a factor of $(\Delta T)^2$ with increasing temperature[b] from T = 642-646 °C. Extrapolating this trend to the lower temperature range shows a nearly imperceptible variance from thermal motion (≤ 125-nm) that is obscured by the dislocation interactions (shown in the Supplement). We reference the thermal motion based on the mean spacing between adjacent dislocations in the DB in Fig. 5a. Comparison between the two curves reveals that by 646 °C, the variance in the position of each dislocation is higher than the mean spacing between dislocations. Fig. 5b gives a closer look at the dislocation spacing in the DB, plotting histograms for each temperature with colored bars that show the contributions from each D—D pair. The black lines and text show the normal distribution and mean spacing for each temperature, matching Fig. 5a. Contributions from each dislocation pair reveal that the increased variance arises from a widening spread in the spacings between each dislocation pair, showing that the boundary becomes more inhomogeneous with each temperature increase.

Dislocation theory describes thermally-activated dislocation motion as high-temperature creep, which gradual increases dislocation mobility with temperature (*36*), however, the data in Fig. 5a-b shows a more severe effect. The positional variance of each dislocation increases beyond the average spacing between dislocations as the boundary spacing becomes heterogeneous, demonstrating effects beyond classical plasticity theory.

[b] This corresponds to mean variances at T = 642, 644, and 646 °C of 0.5, 2, and 8, respectively.

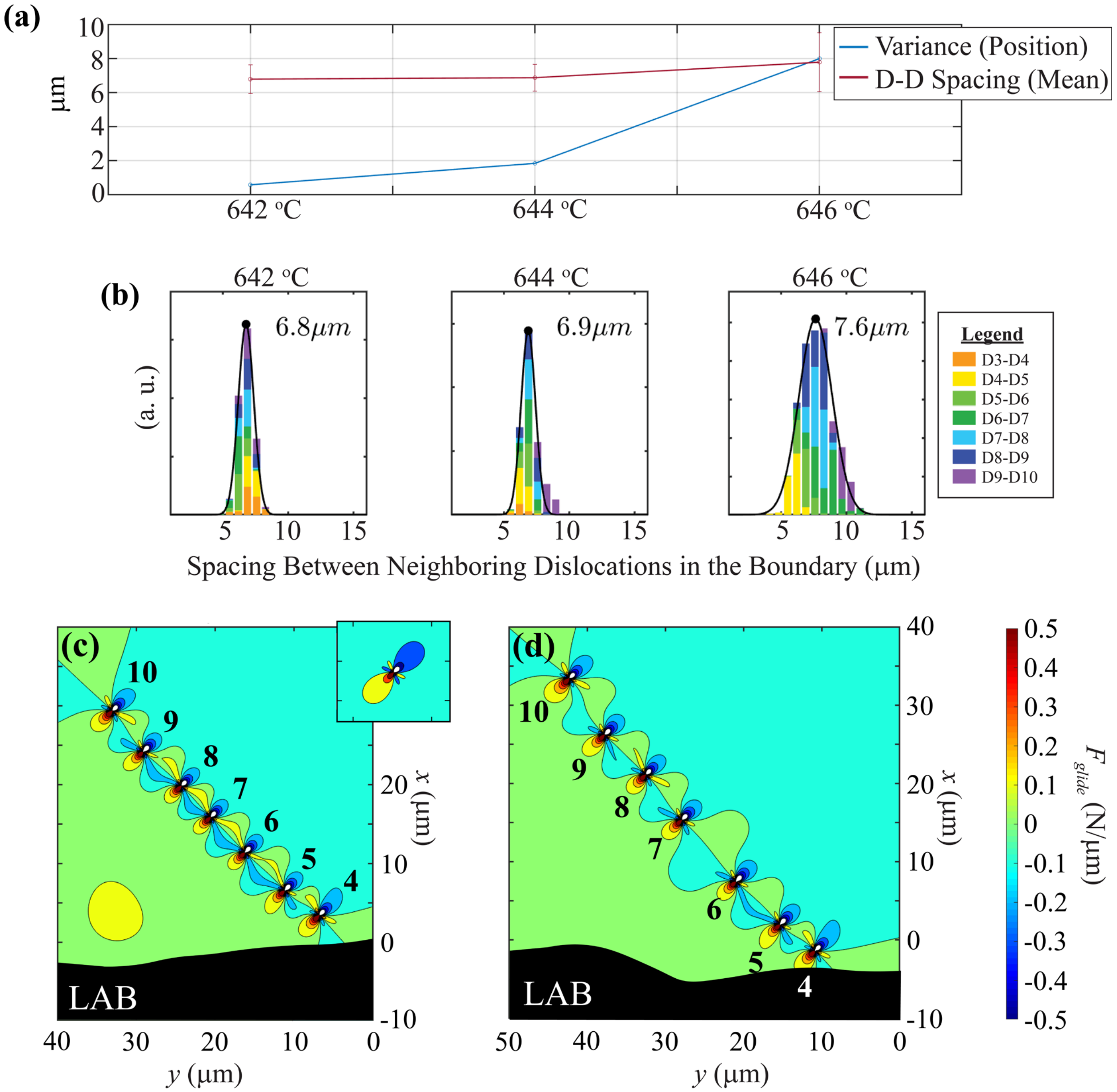

**Figure 5.** A view of the motion of dislocations in the boundary that show how it destabilizes. (a) Plot showing the variance in the time-averaged position (measured as the distance of each DB dislocation from the (0,0) position), plotted at the highest temperatures (blue). This is overlaid with a plot of the average spacing between neighboring dislocations at each temperature, with errorbars to show the variance. Both plots are collected based on the average from all 7 dislocations in all 500 frames for each movie (~120-s acquisition). (b) Histograms showing the distribution of the spacing between dislocations in the DB as the boundary destabilizes, with colors in the bars to show contributions from each pair of dislocations (colors are referenced to the higher dislocation of each pair). We plot the normal distribution fitted from all dislocation pairs in black and label the mean spacing for each temperature. (c-d) Force field calculations along the glide direction, simulated based on the positions of each boundary dislocation identified in Fig. 4, to show changes to the boundary stability. Traces are computed based on frames (c) shortly after the temperature jump at T = 642 °C, and (d) at T = 646 °C. The inset in (c) shows a lone dislocation. For ease of comparison, all force fields are plotted with the same scaling and contoured to the same thresholds.

The dislocation densities reported here are substantially lower than those of most conventional experiments, allowing us to study the interactions in a DB without interaction forces from competing dislocations. The wide spacing between neighboring dislocations causes their interaction forces to be low (~0.5 N/μm), resulting in minimal stress screening. This indicates that under these unique conditions, the boundary is likely only stabilized because of the absence of competing interactions due to the very low dislocation density. Fig. 5c maps the simulated interaction forces along the glide direction that stabilize the DB for the dislocations we identify in Fig. 4 at T = 642 °C. We only predict forces along the glide direction, as the corresponding climb interaction forces are significantly weaker than the ones arising from vacancy diffusion above ~$0.9T_m$ (full discussion of force calculations in Supplement). At T = 642 °C, the variance is quite low and the spacing between dislocations is relatively uniform; the long-range stress fields for each dislocation are thus screened by their interaction forces along the DB[c] (*6*). The screening is clearest when comparing the size and magnitude of the stress fields in the DB to those produced by a lone dislocation, as shown in the inset of Fig. 5c.

In contrast, at only 4 °C higher temperature, the wider and inhomogeneous spacing in the DB no longer screens each DB dislocation uniformly. Fig. 5d plots with the force fields at the time just before D4 and D5 absorb into the LAB together. With this non-uniform spacing, some dislocations group into smaller segments—analogous to the initially separate segments before the DB coalesces (Fig. 2b). As dislocations drift towards each other along the boundary plane, they leave wide spaces in the boundary, as in the case of D6&7 and D9&10, which leave weakened links along the chain. While this might suggest the boundary should break into smaller units, the dynamics in Fig. 4 reveal that thermal motion allows dislocations to group and re-group in different arrangements (see Supplementary Movie 2). The successive re-groupings explain why D6 does not absorb into the LAB when D4&5 exit; at the time of absorption, D6 shifts back towards D7 as non-boundary dislocations interact to catalyze the absorption. We thus observe dislocations exiting the boundary only at its ends, where interactions with other dislocations and the LAB localize.

The massive increase in the disordered motion and spacing as DB dislocations drift apart at these high temperatures reveals the mechanism by which dislocation boundaries natively destabilize on the verge of melting. The uniform spacing in these types of DBs ultimately gives rise to their inherent stability. Only 8 °C earlier, this DB stability causes two DB segments to coalesce (Fig. 3), however, at $0.99T_m$ the boundary dislocations appear to cluster into random groupings. By separating into smaller groups, the boundary loses much of the stability it gains from the long dislocation chain, reducing the stress screening effect and eventually canceling out the inherent stability of the boundary. These destabilizing effects likely explain why *two* dislocations get absorbed into the LAB at only this temperature, corresponding to D4 and D5 in Fig. 5c-d (positions just before absorption).

While TEM has resolved DB dissolution driven by migration at higher dislocation densities (*37*), our conditions and long-range interaction forces present a new view of a DB unraveling from only its internal forces. This statistical view of the stochastic dislocation motion allows us to now quantify a lower-bound for the force necessary to stabilize an isolated tilt boundary before it inhomogeneously breaks apart. Our findings demonstrate just one example of the exotic behavior of dislocation structures at the conditions relevant for melting. At this time, a lack of experimental data on the dynamics of sub-surface dislocations as $T \rightarrow T_m$ has limited melt theory to simplified systems that under-sample the physics at high-T. New computational approaches have demonstrated that just as the symmetry of the lattice is about to transform into a disordered liquid, a plethora of defect interactions create numerous degenerate pathways for melting (*38*). Our new results and approach present an important step forward in directly resolving the relevant physics required to build a valid model that describes melting.

---

[c] Interaction forces between dislocations are described by $F = \sigma b$ for each dislocation, where $F$ is the force, $\sigma$ is stress and $b$ is the Burgers vector. Thus, interaction forces scale linearly with stress.

**Conclusions**

In this first study with time-resolved DFXM, we demonstrate key new insight into collective dislocation interactions at temperatures on the verge of melting, resolving length- and time-scales not previously accessible in a setting representative of bulk behavior. While a static view of the DB's evolution demonstrates that dislocations come and go as the dislocations spread apart, a time-resolved view of each component dislocation over the full 12.5-minute scan reveals how these phenomena occur. We resolve the mechanism by which two DB segments coalesce to stabilize the full structure, then extent this to view how successive dislocation absorption and escape events reduce the number of dislocations in the boundary. At the highest temperatures, we reveal how stochastic thermal motion as the dislocations migrate apart destabilizes the boundary at temperatures approaching the melt. With force-field simulations, we demonstrate that the weakened interaction forces at the highest spacings begin to compete with thermal motion—mobilizing dislocations to cluster and re-cluster into different segments. Our comparison to elastic theory demonstrates how the spacing and inhomogeneity reduce the stabilizing forces, causing the DB to begin to unravel at 0.99 $T_m$. Linking our microscopic pathways to the dynamics of the material in this way enables us to resolve key multi-scale dislocation interactions that models struggle to predict, in the context of melting, deformation, geophysics and beyond. While current multi-scale experiments typically connect the microstructure to bulk properties with only the dislocation density, our study with time-resolved DFXM reveals how new details about the dynamics provide important metrics to quantify the evolution of dislocation structures and connect them to the bulk. Our new approach is an important step forward to connect defects to macroscopic properties.

**Acknowledgments:** We acknowledge Ibo Matthews, Luis Zeppeda-Ruiz, Amit Samanta, W. Craig Carter and Ricardo Pablo-Pedro for their helpful discussions about this work. We thank Ragnvald Mathiesen and the ESRF for allocation of beamtime.

**Funding:** This work was performed under the auspices of the U.S. Department of Energy by Lawrence Livermore National Laboratory under Contract DE-AC52-07NA27344. We also acknowledge the Lawrence Fellowship, which funded much of the work in this work. GW and HFP acknowledge support from the ESS lighthouse on hard materials in 3D, SOLID, funded by the Danish Agency for Science and Higher Education, grant number 8144-00002B, and a grant from the European Research Council "The Physics of Metal Plasticity", grant number ERC-2019-ADV-885022. This manuscript has been authored in part by Mission Support and Test Services, LLC, under Contract No. DE-NA0003624 with the U.S. Department of Energy and supported by the Site-Directed Research and Development Program, U.S. Department of Energy, National Nuclear Security Administration. The United States Government retains and the publisher, by accepting the article for publication, acknowledges that the United States Government retains a non-exclusive, paid-up, irrevocable, worldwide license to publish or reproduce the published form of this manuscript, or allow others to do so, for United States Government purposes. The U.S. Department of Energy will provide public access to these results of federally sponsored research in accordance with the DOE Public Access Plan (http://energy.gov/downloads/doe-public-access-plan). The views expressed in the article do not necessarily represent the views of the U.S. Department of Energy or the United States Government. DOE/NV/03624--0762. We also thank Danscatt for a travel grant.

**Author contributions:** LEDM, PK, CY, HS, MK and CD performed the experiment. Data from the experiments was analyzed by LEDM, MH, AG and SRB, and complementary simulations and interpretation were carried out by LEDM, GW and HFP. The manuscript was written primarily by LEDM, GW, HFP and JE with revisions from all co-authors.

**Competing interests:** All authors declare no competing interests.

**Data and materials availability:** All data is available in the main text or the supplementary materials.

**Supplementary Materials:**

# Supplementary Materials for

## *In-situ* Visualization of Long-Range Defect Interactions at the Edge of Melting

Leora E. Dresselhaus-Marais[1*], Grethe Winther[2], Marylesa Howard[3], Arnulfo Gonzalez[3], Sean R. Breckling[3], Can Yildirim[4], Philip K. Cook[5], Mustafacan Kutsal[2,6], Hugh Simons[7], Carsten Detlefs[6], Jon H. Eggert[1], Henning Friis Poulsen[7]

Correspondence to: dresselhausc1@llnl.gov

**This PDF file includes:**

**Other Supplementary Materials for this manuscript include the following:**

**Materials and Methods**

This experiment was performed at the dedicated dark-field X-ray microscope instrument at ID-06 at the European Synchrotron Radiation Facility (ESRF) *(39)*, with the instrument placed 57 m from the source. A Si(111) Bragg-Bragg monochromator defined a 17.29 keV X-ray beam. The beam was collimated in the vertical direction by a compound refractive lens (CRLs), comprising 8 Be 2D CRLs with a radius of curvature of R = 200 μm. Then it was focused by a condenser comprising 55 Be 1D lenses with R = 100 μm corresponding to an effective aperture of 435 μm, a focal length of 816 mm, a divergence of $\zeta_v = 0.030°$ (FWHM) and a nominal focal spot height of 220 nm. After illuminating the sample, the {002} diffracted beam was magnified by an X-ray objective (a CRL with 88 2D Be lenses with R = 50μm, T = 2 mm) positioned 274-mm behind the sample, and a far-field CCD camera placed 5.364 m from the sample. The resulting numerical aperture was 0.720 mrad (RMS), with a measured magnification of 18.5x. The 2D far-field detector used a scintillator screen coupled to a FreLon CCD camera via microscope optics, giving an effective pixel size (in the sample plane) of 75 nm/pixel along the *y* direction and 204 nm/pixel along the *x* direction. We used an optical furnace to evenly heat our sample, as described elsewhere *(40)*.

The sample was a 0.5 x 0.5 x 20 $mm^3$ single crystal of aluminium with 6N commercial purity, used as purchased from Surface Preparation Laboratory. The mosaic spread of the single crystal was within ~0.2° across our 400-μm region of interest. This spread reduced significantly as the crystal was incrementally annealed in the experiment, following the temperature path detailed in Supplementary Text S4. The instrument was aligned such that it could image the small variations in reciprocal space around the nominal (002) diffraction vector *(41,42)*, with the associated scattering angle $2\theta = 20.73°$. Similar to the approach presented in Jakobsen *et al. (31)* the sample tilt was slightly offset from the peak of the rocking curve, to create the conditions for weak-beam contrast in dark-field microscopy. The spatial resolution is estimated from the sharpest feature in the image to be ~300 nm in the sample plane. The FWHM of our reciprocal-space resolution were defined by the NA of the lens: $\Delta 2\theta = 0.134°$ and $\Delta\eta = 0.134°$, respectively *(41)*. The primary reciprocal-space contrast in this experiment arises from the divergence of the incoming beam (in the rocking direction), $\Delta\chi = 0.015°$ *(31,41)*. (For definition of angles, see Supplementary Fig. S1) A full description of the contrast mechanisms in our experimental geometry for DFXM are described in full elsewhere *(31)*.

We used a radiation furnace to evenly heat our sample, as described in Yildirim, et al. *(40)*. Initially the sample was heated from room temperature to 633 °C over ~10 hours. The experiments presented in this work focused on the end of this heating path, during which the temperature was increased by ~2 °C increments (in ~5 seconds) followed by realignment and subsequent data acquisition for 2-5 minutes. Each image was collected over 100-ms integration times, acquired at the 4-Hz frame rate of the camera, with all elements of the microscope and sample remaining stationary. All images were collected with the motor positions and background subtractions from data collected during the experiment. No subsequent normalization was performed.

## Supplementary Text

### S1—Interpretation of Burgers Vectors

We interpret the Burgers vectors and slip planes for the dislocations in the DB based on their motion, packing arrangement and spacing.

We compared the dislocation motion in our experiments to glide and climb motion for each of the 12 possible Burgers vector and slip plane combinations possible for edge dislocations in FCC crystals. The directions of motion we observe are consistent with Burgers vectors of either ½$[110]$ or ½$[1\bar{1}0]$. The [1,-1] trace of the DB matches the stable plane for a tilt boundary of ½$[1\bar{1}0]$ dislocations. We distinguish between the two based on the trace of the DB in the main text. In addition, our assignment of the ½$[1\bar{1}0]$ matches with interpreting the relatively smooth collective motion of the DB as glide, whereas the less smooth motion of the dislocations along the DB trace agrees with climb requiring vacancy diffusion.

Our experiment has focused on time-resolving the dislocation dynamics in aluminum—a well-studied prototypical material to study dislocation interactions. As some boundary conditions are lacking (the study only reveals the dynamics in one plane), we refrain from comparing to full dislocation dynamics simulation and instead focus on developing the data analysis approach to study the dynamics at these large length-scales. We have therefore focused our interpretation of the Burgers vector around the directions of dislocation motion and the stabilizing effects that we observe in the DB structure. The observed orthogonal directions of dislocation motion in the present geometry only match the glide and climb directions of dislocations of $\vec{\boldsymbol{b}}$ = $[1\bar{1}0]$ or $\vec{\boldsymbol{b}}$ = $[110]$. It is known that the only stable boundary with Burgers vectors of one type has a boundary plane that is orthogonal to the Burgers vector. For $\vec{\boldsymbol{b}}$ = $[1\bar{1}0]$, the trace of the boundary plane would produce the observed 45° inclination in our observation plane and is orthogonal to that. In this configuration the dislocations screen each other's long-range strain fields. Dislocations of $\vec{\boldsymbol{b}}$ = $[110]$ would form a tilt boundary with a trace of -45°. The interpretation of the Burgers vector as $\vec{\boldsymbol{b}}$ = $[1\bar{1}0]$ also matches the expectation that the fairly smooth motion of the entire DB upon temperature changes occurs by glide, whereas the less smooth motion of the dislocations in the DB is due to the fact that climb requires diffusion of vacancies. At this point we cannot identify whether the slip plane is $(111)$ or $(11\bar{1})$.

We note that we see contrast for this $\vec{b}$ = $\langle 1\bar{1}0 \rangle$ dislocation along the $\vec{g} = \langle 002 \rangle$ diffraction peak because the line vector $l$ = $\langle 112 \rangle$ causes the geometry to satisfy the condition for sensitivity $g \cdot b \times l \geq 0.64$, as in dark-field TEM *(43)*.

### S2—Full Mapping of Dislocation Features

The geometry of the hard X-ray microscope in general is presented by Poulsen in previous work *(41-42)*. We sketch the geometry of our experiment in Fig. S1(a), with a coordinate system for the laboratory and another that is fixed to the main crystallographic axes; rotation of the crystal thus rotates the coordinate systems with respect to each other by an angle, μ, around the positive *x*-axis. For ease of presentation in the main text we have neglected this angle of μ = 10° and set the observation plane in direct space to be perpendicular to the diffraction vector **g**. As we are not attempting direct quantitative comparisons between experimental data and model that is permissible.

To improve time resolution in this work we mainly relied on acquiring movies at a fixed ($2\theta$, $\mu$, $\eta$). As discussed in detail in established work on DFXM *(28, 31, 41)*, one can acquire a full map of selected strain components along a reciprocal lattice vector by scanning the sample around $\mu$ (a rocking scan) and by combined movements of the objective and detector in the $2\theta$ and $\eta$ directions. To provide an overall impression of the topology of the sample, a ($2\theta$-$\mu$) map was collected at 606 °C, which is reproduced in Fig. S1(b). As the collection of one ($2\theta$-$\mu$) map requires ~30 minutes, these detailed scans were not attainable for T > 606 °C, as the dislocation structures were not stationary over the necessary acquisition time.

## S3—Details of Fitting for DB Snapshots

Fig. 2 in the main text summarizes the positions and motion of the DB over this temperature range. We provide maps of each images used to compile Fig. 2 in Fig. S2, showing the position and angle of the dislocation boundary at each temperature. The position and tilt of all dislocation boundary segments were found by inputting the midpoints of each dislocation feature into standard linear fitting and regression methods. The resulting fit equations are given in Table S1.

## S4—Thermal History of Sample

We use an optical furnace in this work, which illuminates the sample with blackbody infrared radiation (IR) to heat the sample, as described and characterized fully elsewhere *(41)*. As Yildirim demonstrates in that work, the heat load uniformly covers a ~1 x 1 x 1 $mm^3$ region of the sample, heating continuously with electrical current in the IR source. At temperatures for $T < 0.94\ T_m$, we did not observe changes between each frame collected at the maximum repetition rate of 4 Hz (limited by photon flux and data transfer rates). We therefore calibrated the relationship between DC electric power (DEPow) and temperature from the carefully aligned data collected at these temperatures, using the thermal shift from the diffracted beam to measure thermal expansion in the lattice constant *(44)*. Fitting the calibrated datasets to a line, we then computed the corresponding temperatures at $T \geq 0.94\ T_m$ by computing each DEPow from this fit line (Fig. S4). By fitting the deviations from the fitted data with a Gaussian distribution, we estimate that this adds ± 0.1% uncertainty to the temperatures given in this work. We verified that these measured temperatures are consistent with the melting temperature at 660 °C.

## S5—Details for Simulations of Interaction Force-Fields

The calculations in this work were limited to applications of elastic theory, as mentioned above *(6)*. For these calculations, we transformed the dislocation positions into a new coordinate system that aligns the relevant Burgers vector, line vector and slip-plane normal with the coordinate axes, following the assignments detailed in Supplementary Section S1. In this system, $x$ defines the glide direction, $y$ defines the climb direction, and $z$ defines the direction of the dislocation line. Force-field maps in Fig. 5c-d of the main text were compiled based on the interaction forces between nearby infinite straight dislocation lines, as described in Chapter 4 of *(6)*. We define each dislocation's position from our experimental data and compute their force fields by summing contributions from each dislocation as

$$F_x = \sigma_{xy} b = \frac{Gb^2}{2\pi(1-\nu)} \left( \frac{x(x^2-y^2)}{(x^2+y^2)^2} \right).$$

We do not provide the analogous simulations for the climb direction, as climb motion is dominated by vacancy diffusion. Vacancy diffusion forces have higher magnitudes at 0.97-0.99 $T_m$ and pull in random directions, which is consistent with the random motion we observe.

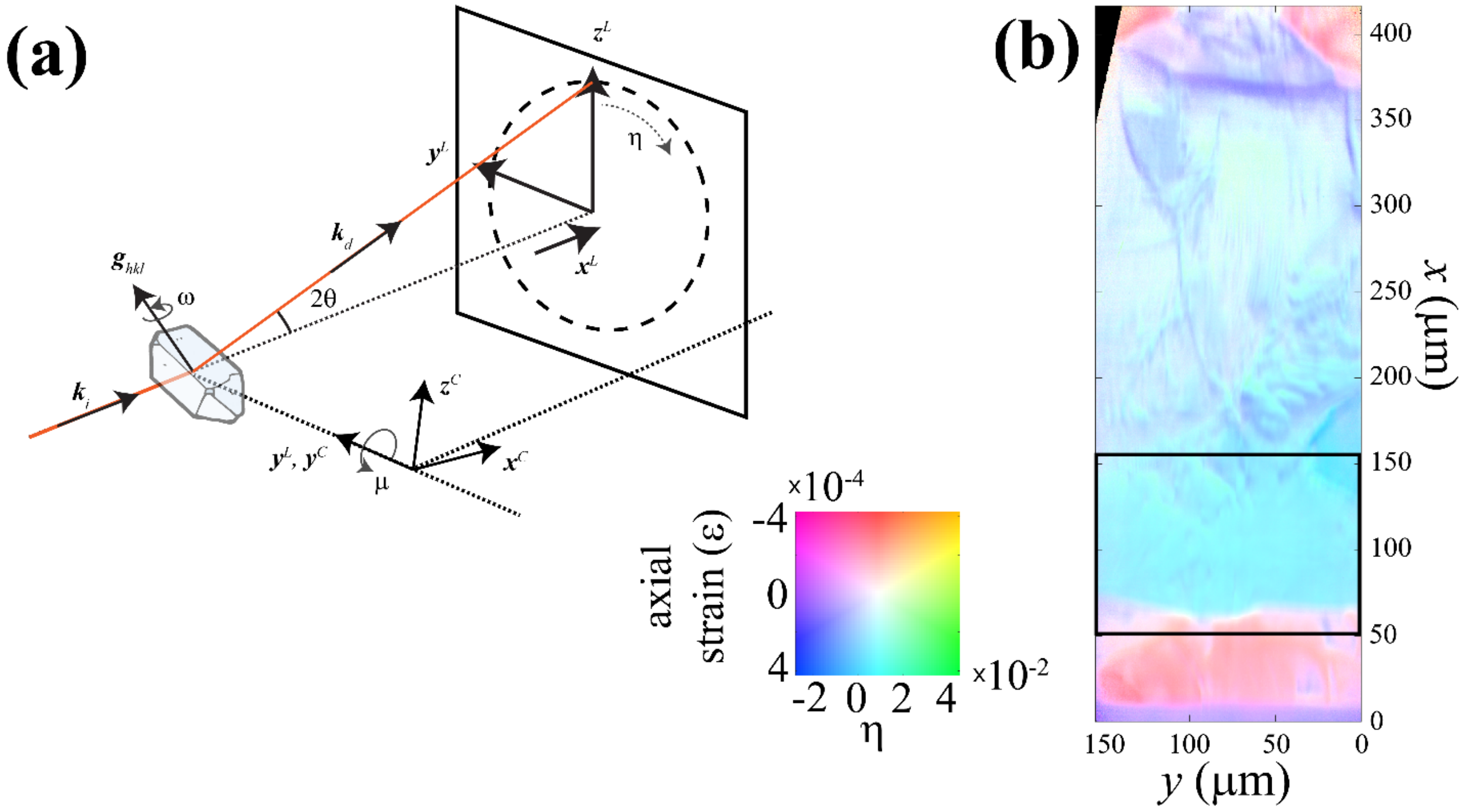

**Fig. S1.**

**(a)** A schematic of the diffraction geometry that we used in our experiments, with all axes defined in the laboratory frame (*L*) and the crystallographic frame (*C*) of reference. Incident and diffracted wavevectors are labeled as $\boldsymbol{k}_i$, and $\boldsymbol{k}_d$, respectively, and the reciprocal lattice vector is labeled as $\boldsymbol{g}_{hkl}$. We also indicate rotation around the Debye-Scherrer ring as η and rotation of the crystal about the *y*-axis as μ. **(b)** Full axial strain and rotation map of the same single crystal of aluminum that was used in the main text, with all immobile (sessile) dislocations at T = 606 °C at the length-scale presented in this work. The black boxed region over the image shows the sub-section of the crystal that is displayed in Fig. 1b. The LAB discussed in the main text corresponds to the region where the color changes from pink to blue at the top of the black boxed region, with its ~0.01° misorientation.

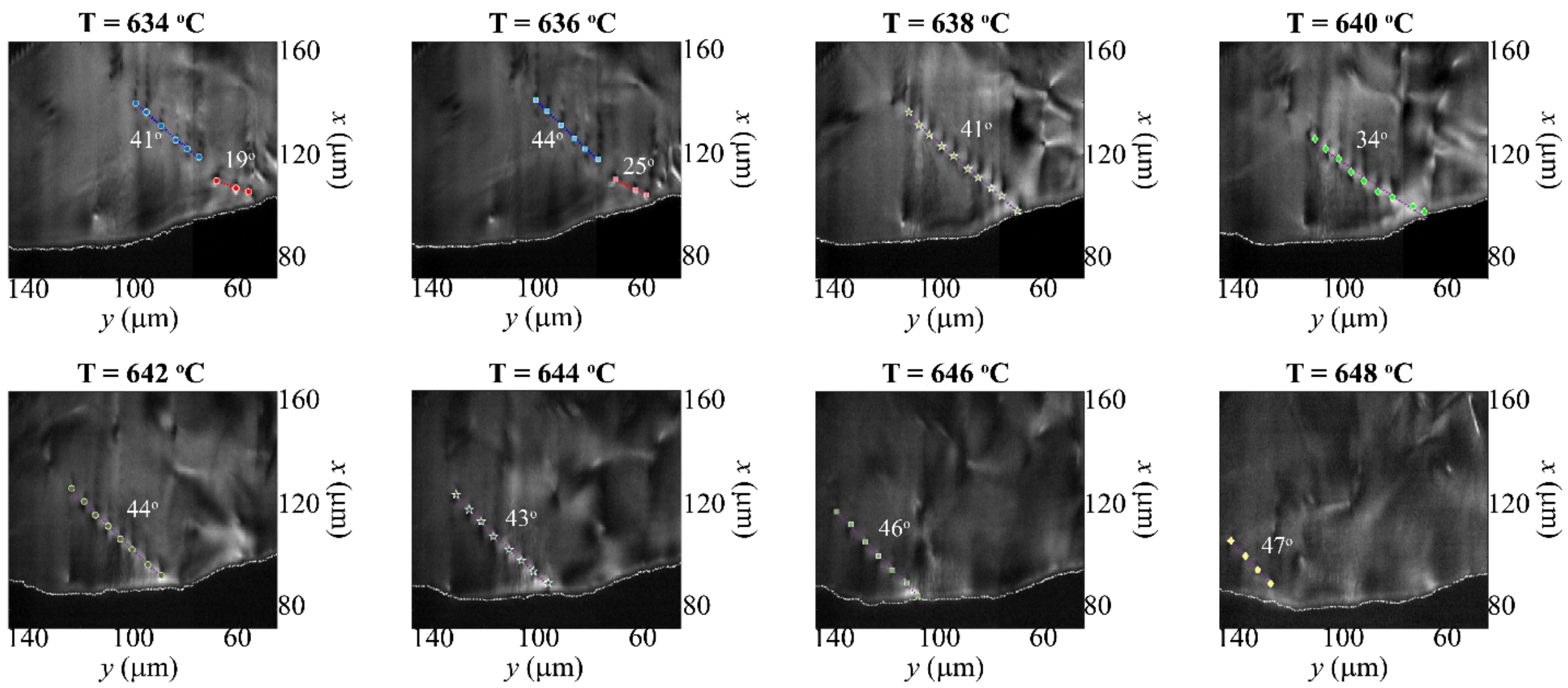

**Fig. S2.**

Changes to the position and angle of the DB shown in Fig. 1(b) of the main text at the 100$^{th}$ frame for each temperature. We compile each DB segment into a line to define the angle between the trace of the DB packing plane and the horizontal axis. The LAB is highlighted by a thin dotted white line in all frames.

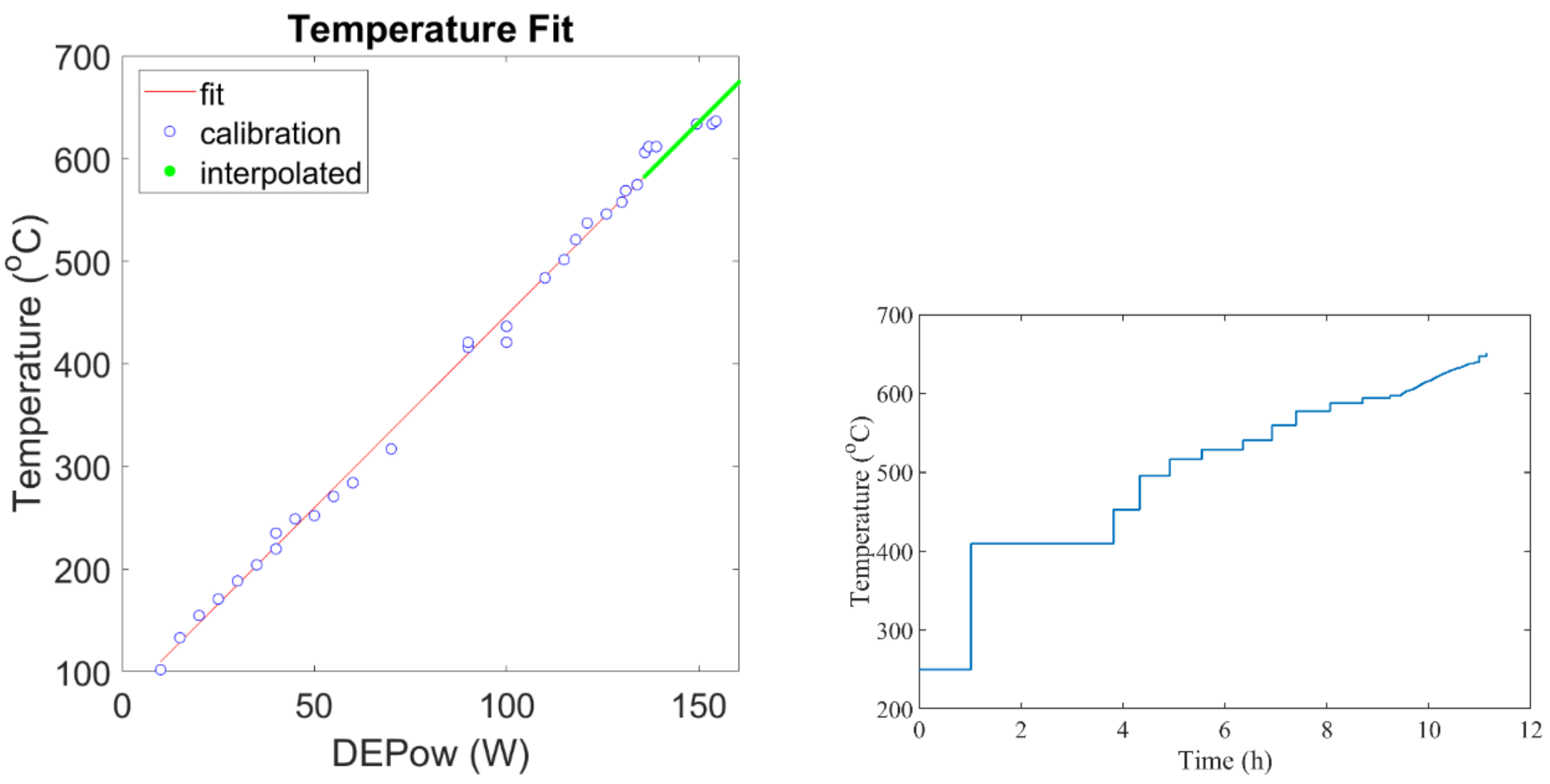

**Fig. S3.**

(left) Plot of the temperature calibration curve between the DC power in the optical furnace and the temperature of the aluminum crystal, as computed based on thermal expansion of the lattice constants. (right) Temperature history path of the sample from the beginning of the experiment through the melting transition. The measurements reported in this work were conducted beginning at $t$ ~10.5 hours.

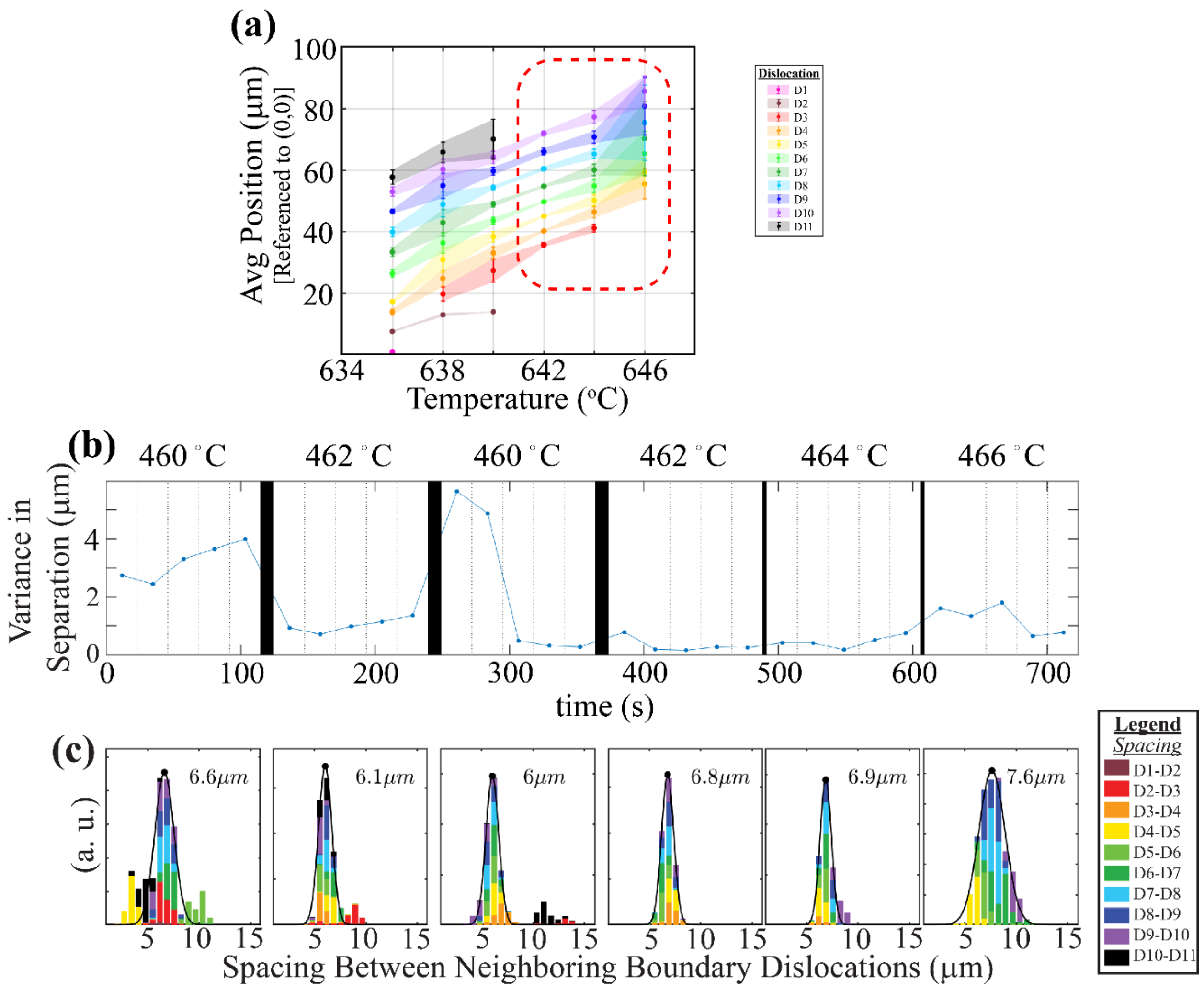

**Fig. S4.**

(a) Plot of the full time-averaged position and variance for each dislocation at all temperatures considered in this work. The trends discussed in Fig. 5 of the main text are circled in the red dashed line. Trends at lower temperatures show high variance based on dislocation interactions that are discussed in the main text. (b) Plot of the variance in spacing between adjacent dislocations over increments of ~25 s (100 frames), showing trends that the dislocations are most mobile as dislocations leave the boundary. (c) Full temperature progression of the histograms shown in Fig. 5 of the main text, demonstrating that the low-temperature variance is dominated by dislocation interactions.

**Table S1.**
Fit lines compiled with regression analysis, which are used to define the DB and its trace angle at each temperature in Fig. S2.

| **Temperature** | **Linear Fit** | **Angle** | **R2** |
|---|---|---|---|
| T = 634 °C *(Upper Segment)* | $x = 0.33773y + 86.8898$ | 18.6616° | 0.99247 |
| T = 634 °C *(Lower Segment)* | $x = 0.8661y + 53.4863$ | 40.8957° | 0.99755 |
| T = 636 °C *(Upper Segment)* | $x = 0.4702y + 77.3756$ | 25.183 | 1 |
| T = 636 °C *(Lower Segment)* | $x = 0.96085y + 44.1219$ | 43.8561° | 0.99899 |
| T = 638 °C *(Full DB)* | $x = 0.88161y + 36.1144$ | 41.3996° | 0.9975 |
| T = 640 °C *(Full DB)* | $x = 0.69076y + 48.0271$ | 34.6351° | 0.97646 |
| T = 642 °C *(Full DB)* | $x = 0.97203y + 5.2261$ | 44.1875° | 0.99784 |
| T = 644 °C *(Full DB)* | $x = 0.95572y + -2.3423$ | 43.7031° | 0.99579 |
| T = 646 °C *(Full DB)* | $x = 1.0361y + -27.484$ | 46.0152° | 0.99888 |
| T = 648 °C *(Full DB)* | $x = 1.0761y + -48.0238$ | 47.0986° | 0.99978 |

**Movie S1.**

Real-time movie of D3 inserting into the lower segment of the DB at T = 638 °C.

**Movie S2.**

Real-time movie of the three highest temperatures, showing dislocation motion from T = 642-646 °C, as plotted in Fig. 4-5 of the main manuscript. Circles are drawn to show the position of each dislocation, following the color system in Fig. 4. The crystal orientation is drawn and a reference position (0,0) is marked with a star in all frames. Each temperature step is shown with blank frames, separated at the same 4-Hz repetition rate to span the heating time.